\documentclass[12pt,preprint]{aastex}

\shorttitle{IRC+10216 CS(14--13)}
\shortauthors{Young et al.}

\begin{document}

\slugcomment{Accepted by ApJ Letters: March 8, 2004}

\title{Submillimeter Array Observations of 
CS J=14--13 Emission from the Evolved Star IRC+10216}

\author{K.H. Young\altaffilmark{1},
        T.R. Hunter\altaffilmark{1},
        D.J. Wilner\altaffilmark{1},
        M.A. Gurwell\altaffilmark{1}, \\
        J.W. Barrett\altaffilmark{1},
        R. Blundell\altaffilmark{1},
        R. Christensen\altaffilmark{2},
        D. Fong\altaffilmark{2},
        N. Hirano\altaffilmark{3}, 
        P.T.P. Ho\altaffilmark{1,3},
        S.Y. Liu\altaffilmark{3}, 
        K.Y. Lo\altaffilmark{3,5},
        R. Martin\altaffilmark{3}, 
        S. Matsushita\altaffilmark{3}, 
        J.M. Moran\altaffilmark{1},
        N. Ohashi\altaffilmark{3},
        D.C. Papa\altaffilmark{1},
        N. Patel\altaffilmark{1},
        F. Patt\altaffilmark{3}, 
        A. Peck\altaffilmark{2},
        C. Qi\altaffilmark{1},
        M. Saito\altaffilmark{1,4}, 
        A. Schinckel\altaffilmark{2},
        H. Shinnaga\altaffilmark{2},
        T.K. Sridharan\altaffilmark{1},
        S. Takakuwa\altaffilmark{2},
        C.E. Tong\altaffilmark{1},
        D.V. Trung\altaffilmark{3} 
}
\email{dwilner@cfa.harvard.edu}
\altaffiltext{1}{Harvard-Smithsonian Center for Astrophysics, 
  60 Garden Street, Cambridge, MA 02138, USA}
\altaffiltext{2}{Harvard-Smithsonian Center for Astrophysics, 
   Submillimeter Array, 645 N. A'ohoku Place, Hilo, HI 96721, USA}
\altaffiltext{3}{Academia Sinica Institute of Astronomy \&
   Astrophysics, P.O. Box 23-141, Taipei, Taiwan, 106, R.O.C.}
\altaffiltext{4}{National Astronomical Observatory of Japan,
2-21-1 Osawa, Mitaka, Tokyo 181-8588, Japan}
\altaffiltext{5}{National Radio Astronomy Observatory, 
520 Edgemont Road, Charlottesville, VA 22903, USA}

\begin{abstract}
We present imaging observations of the evolved star IRC+10216 
in the CS J=14--13 line at 685.4 GHz and associated submillimeter 
continuum at $\sim 2''$ resolution made with the partially constructed 
Submillimeter Array. 
The CS J=14--13 line emission from the stellar envelope is
well resolved both spatially and spectrally. 
The strong central concentration of the line emission provides 
direct evidence that CS is a parent molecule that forms close to 
the stellar photosphere, in accord with previous images of the 
lower excitation CS J=2--1 line and inferences from unresolved
observations of vibrationally excited transitions.
The continuum emission is dominated by a compact, unresolved 
component, consistent with the photospheric emission,
that accounts for $\sim20\%$ of the broadband 450~$\mu$m flux.
These are the first interferometer imaging observations 
made in the semi-transparent 450~$\mu$m atmospheric window.
\end{abstract}


\section{Introduction}
The nearby carbon star IRC+10216 (CW Leo, IRAS 09452+1330) is surrounded 
by a dusty, expanding envelope with relatively simple geometry and kinematics 
that has been the subject of many studies of astrochemical processes.
A rich chemistry develops in the envelope as gas is lost from the 
star at a high rate ($\sim3\times10^{-5}$~M$_{\odot}$~yr$^{-1}$,
Crosas \& Menten 1997) and is modified by thermodynamic equilibrium and 
non-equilibrium reactions, photochemical reactions, ion-molecule reactions, 
and the condensation of dust grains. 
The range of physical conditions in the envelope make it especially 
well suited to observations of rotational emission lines from molecules, 
and numerous species have been identified in spectral scans at millimeter 
and submillimeter wavelengths 
(e.g. Cernicharo, Guelin \&  Kahane 2000; Groesbeck, Phillips \& Blake 1994,
Avery et al. 1992).

The close proximity ($\sim150$~pc) and copious mass loss of IRC+10216 allows 
studies of the circumstellar envelope on a wide range of size scales.
On the largest scales, an extensive envelope of dust 
(Mauron \& Huggins 1999) and gas (Fong, Meixner \& Shah 2003) has been 
found to extend to $\sim0.15$~pc ($\sim3\farcm3$) in radius.  
On much smaller scales,
interferometric imaging at millimeter wavelengths has revealed the spatial 
distribution of molecular species within the central arcminute
(e.g. Bieging \& Tafalla 1993,  
Guelin, Lucas \& Cernicharo 1993,
Lucas \& Guelin 1999).
Parent molecules are found to be centrally peaked (e.g. SiS, HCN), while 
daughter molecules are found with shell distributions 
(e.g. CN, C$_2$H, HNC). High angular resolution is especially important for 
imaging lines of high excitation that emerge from the innermost regions of 
the envelope, where the temperatures and densities are highest. 

Since the high excitation rotational lines of many common molecules 
lie at short submillimeter wavelengths, 
the Submillimeter Array\footnote{ 
The Submillimeter Array is a joint
project between the Smithsonian Astrophysical Observatory and the
Academia Sinica Institute of Astronomy and Astrophysics, and is
funded by the Smithsonian Institution and the Academia Sinica.}
(SMA)
opens new possibilities for probing physical conditions and chemistry 
by enabling imaging observations with arcsecond resolution. 
In this {\em Letter}, we present observations of IRC+10216 in CS J=14--13 
emission made with the partially constructed SMA (Ho, Moran \& Lo 2004).
These are the first interferometer images from observations in the 
atmospheric window centered near 450~$\mu$m, where $\sim$30\% or better 
transmission occurs at high, dry sites like the summit of Mauna Kea a
significant fraction of the time. They build on the pioneering efforts of 
Carlstrom et al. (1994), 
who obtained fringes on the nearby single baseline between the 
Caltech~Submillimeter~Observatory and the James~Clerk~Maxwell~Telescope.
The results provide constraints on sulfur chemistry in the 
inner wind of the IRC+10216 prototype carbon-rich circumstellar envelope.

\section{Observations}

We observed IRC+10216 on 10 December 2002 with the 
SMA when three of the 6~meter diameter antennas were equipped 
with receivers for the ``600--700~GHz'' frequency band. 
The weather was very good, with 225~GHz atmospheric opacity of 
0.035 measured at the nearby Caltech Submillimeter Observatory
through the night,  or $\sim0.7$ at the SMA observing frequency.
Table 1 summarizes the observational parameters. 
We adopt the CS J=14--13 line rest frequency of 685.435923 GHz, 
recently calculated from newly determined 
spectroscopic constants (Gottlieb, Myers \& Thaddeus 2003).
The SMA observations provided three independent baselines ranging
from 14 to 25 meters in length, resulting in $\sim2''$ resolution.

A major difficulty for interferometric imaging at these high
frequencies is the lack of sufficiently strong sources with 
known structure for the derivation of complex gains to track 
instrumental phase drifts and atmospheric fluctuations. In particular, 
nearly all of the point-like quasars 
are too weak for the SMA to detect at shorter wavelengths 
in a few minutes of integration time. 
Compact thermal sources such as planetary
moons or asteroids provide one alternative. For these observations, 
we took advantage of the close proximity of Jupiter to IRC+10216 
in the sky ($<7$~degrees). 
While Jupiter was far too large to be a suitable calibrator 
(diameter $40''$), 
the Galilean moons were only slightly resolved on these baselines,
and high signal-to-noise observations of Callisto (diameter $1\farcs4$) 
could be efficiently interleaved with those of IRC+10216.
Jupiter was located well outside the small SMA field of view 
during the observations of Callisto, and the data show no evidence 
for any contamination from the planetary emission.
IRC+10216 was tracked over the hour angle range $-4$h to $+4$h.
The predicted flux 
of Callisto varied by less than $10$\% due to resolution effects. 

The partially complete digital correlator was configured with four overlapping 
``chunks'', each of 104 MHz bandwidth and 128 channels, with the 
CS J=14--13 line centered in one of them.  A calibration of the bandpass
shape was obtained from observations of Mars. The system temperatures 
ranged from 1200 to 2200 K (DSB).
The absolute flux scale was set assuming a full disk flux of 55~Jy 
for Callisto, based on a brightness temperature of 120 K determined
by comparison with Titan, whose submillimeter spectrum is better known 
(Mark Gurwell, private communication). The overall flux scale is unlikely 
to be more accurate than 30\%.  
Images were made with a variety of weighting schemes that provided
slightly different angular resolutions, sensitivities, and 
synthesized beam sidelobe patterns. 

\section{Results and Discussion}

\subsection{Spectrum at Image Center} 
\begin{figure}
\epsscale{0.53}
\rotatebox{-90}{
\plotone{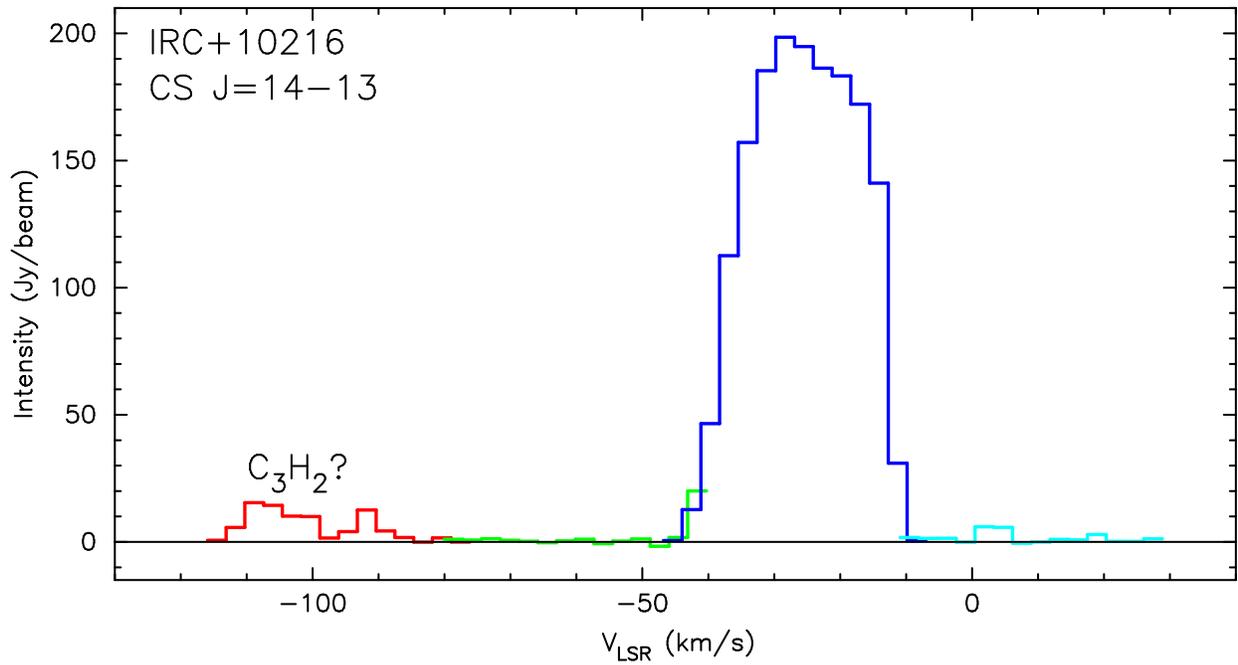}}
\caption{
Spectrum of IRC+10216 across the full bandwidth, 
showing the four overlapping 104 MHz correlator ``chunks'', 
binned to 2.8 km~s$^{-1}$ resolution, 
in a $4\farcs6\times1\farcs6$ p.a. $-68^{\circ}$ synthesized beam. 
Strong CS J=14--13 line emission is visible, as well as an interloping
line at low velocities, tentatively identified as C$_3$H$_2$.
}
\label{fig:cs_spectrum}
\end{figure}

Figure~\ref{fig:cs_spectrum} shows the spectrum from the center of
the image cube for the full USB correlator band, averaging groups of 
8 channels, corresponding to a velocity width of 2.85 km~s$^{-1}$.
The CS J=14--13 line is the prominent feature centered at $-26$~km~s$^{-1}$ 
and an approximate width at zero power of $\sim30$~km~s$^{-1}$, 
values which are consistent with the systemic velocity of the source 
and the expansion velocity of the envelope, respectively.
The line shape is difficult to interpret because spatial filtering 
by the interferometer results in missing flux that varies with 
velocity across the profile (see \S\ref{discussion:line}).
The peak brightness temperature of the line at the 
full angular resolution of the data is well over 100~K. 
An additional feature appears in the spectrum near $-100$~km~s$^{-1}$ 
with a double peaked line shape and an approximate width consistent with 
the circumstellar expansion.  
A search of the JPL spectral line catalog suggests a tentative identification 
of this feature with the C$_3$H$_2$ $8(4,4)-7(3,5)$ line at 685.6125565~GHz.
This species was previously detected in transitions at lower frequencies 
(Kawaguchi et al. 1995, Cernicharo et al. 2000). 
No alternative candidates line were found in the catalog of
Cernicharo et al. (2000) that contains 
1050 molecular species (Cernicharo, private communication).
If the assignment is correct, then the line might be thought 
to emerge from the external shell where various 
carbon chain radicals are found (Guelin et al. 1993). 
This feature appears to be spatially compact, though,
which perhaps renders the identification problematic.

\subsection{Continuum Emission} 
\begin{figure}
\epsscale{0.9}
\rotatebox{-90}{
\plotone{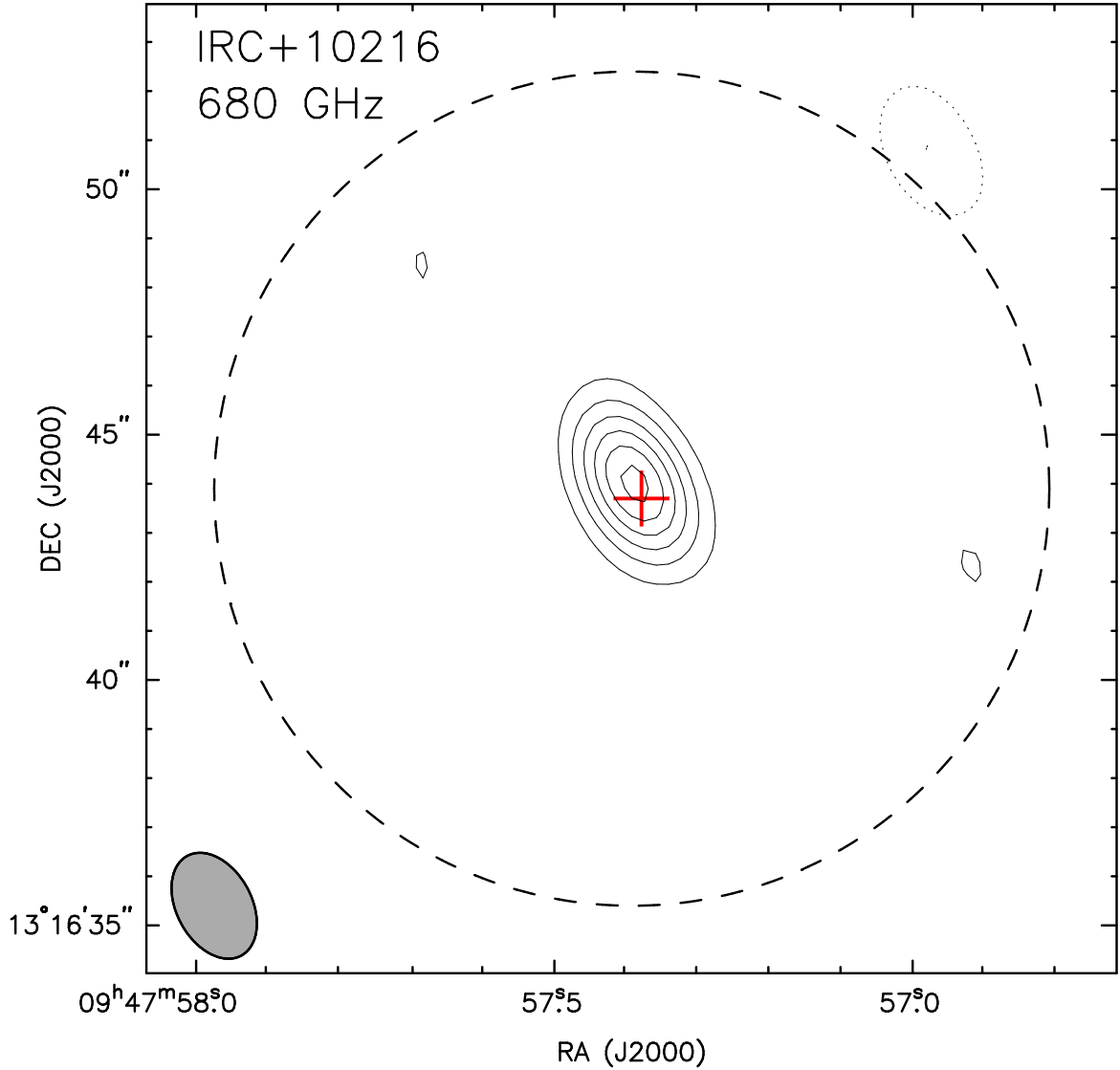}}
\caption{
IRC+10216 continuum emission at 685 GHz.
The contour levels are $\pm2,4,6,...\times0.3$~Jy~beam$^{-1}$.
Negative contours are dotted.
The ellipse in the lower left corner shows the 
$2\farcs3\times1\farcs5$ p.a. $28^{\circ}$ synthesized beam. 
The dashed circle indicates the $17''$ FWHM primary beam size.
}
\label{fig:cont}
\end{figure}

Figure~\ref{fig:cont} shows the continuum image obtained from the 
channels free of strong line emission. Visibilities from both sidebands,
separated by 10~GHz, were combined in a multifrequency synthesis, resulting 
in an effective frequency of 680 GHz. The continuum peak position is 
consistent with previous determinations, in particular the 95~GHz observation 
from the IRAM PdBI (Guelin et al. 1993), which is marked by 
the cross. The peak continuum flux is 3.8~Jy, with an uncertainty 
dominated by systematic effects.

Previous bolometer observations of IRC+10216 at 450~$\mu$m indicate
that the SMA recovers approximately 20\% of the broadband flux at
this wavelength.  Sandell (1994) noted that IRC+10216 has long period 
variations in the submillimeter and found $19\pm3$~Jy in an $18''$ 
aperture near the 635 day maximum.  Jenness et al. (2002) analyzed 
several years of SCUBA observations of IRC+10216 at 850~$\mu$m and 
450~$\mu$m and found a period and maximum flux consistent with the 
earlier single pixel measurements.  The SMA observations were made 
approximately a month before the predicted maximum, when $\sim95\%$ of 
the peak flux would be expected. Remarkably, the CS J=14--13 line emission
{\em alone} accounts for $1\%$ of the flux in the $\sim68$~GHz wide 
filter used for the 450~$\mu$m bolometer measurements. Given the 
many strong spectral lines excited in the inner envelope at these 
high frequencies, it is clear that spectral line emission 
makes an important contribution to the broadband ``continuum'' flux. 

A large fraction of the continuum emission from the compact component 
detected by the SMA at 680~GHz likely comes from the stellar photosphere. 
Lucas \& Guelin (1999) report a ``point source'' component in IRAM PdBI 
observations of $65\pm1$~mJy at 89~GHz 
and $486\pm7$~mJy at 242~GHz, identified as photospheric emission. 
Assuming an optically thick spectrum, $S_\nu\propto\nu^2$, 
an extrapolation to 680~GHz gives 3.8~Jy, consistent with the measured
value.  Thermal emission from dust in the inner envelope cannot contribute 
substantial additional flux in the small synthesized beam of the SMA at 
this high frequency.

\subsection{CS J=14--13 Emission} 
\label{discussion:line}
\begin{figure}
\epsscale{0.66}
\rotatebox{-90}{
\plotone{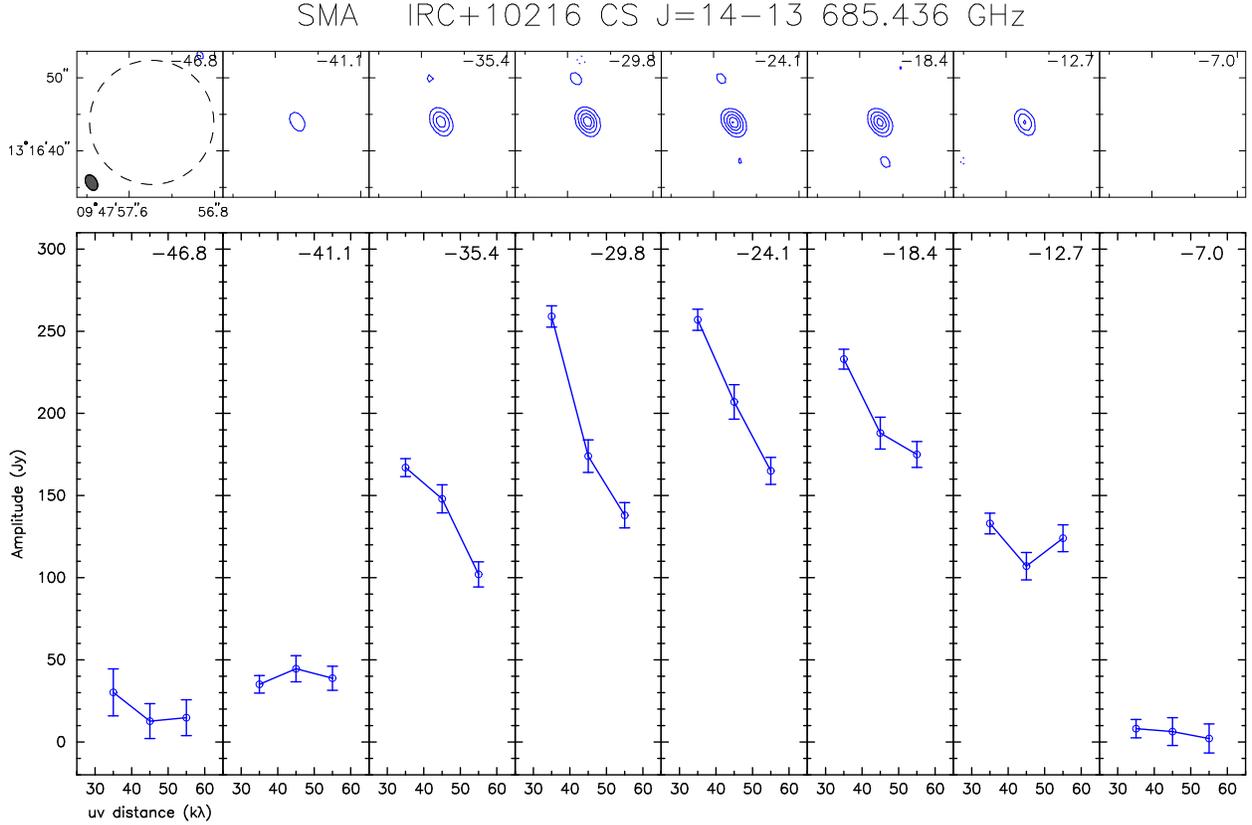}}
\caption{
{\em upper:} Velocity channel maps (width 5.7 km~s$^{-1}$)
of CS J=14--13 line emission from IRC+10216.
The contour levels are $\pm1,3,6,9,...\times15$~Jy~beam$^{-1}$. 
Negative contours are dotted.
The ellipse in the lower left corner of the leftmost panel shows 
the $2\farcs3\times1\farcs5$ p.a. $28^{\circ}$ synthesized beam.
The dashed circle indicates the $17''$ FWHM primary beam size.
{\em lower:}
Visibility amplitude of the CS J=14--13 emission vs. baseline length, 
for the same velocity channels, annularly averaged in 10~k$\lambda$ bins.
The error bars represent $\pm1$ standard deviation for each bin.
The falloff of amplitude with baseline length shows that the emission 
is spatially resolved at the central velocities, as expected for a 
symmetrically expanding source.
}
\label{fig:cs_channels}
\end{figure}

Figure~\ref{fig:cs_channels} shows images for a series of velocity intervals 
that span the CS J=14--13 line. Below each of the images is a plot of
visibility amplitude vs. $(u,v)$ distance for the corresponding velocity 
interval to give an idea of spatial extent, which is not easy
to ascertain from the images.
The overall structure in the visibility amplitudes is consistent 
with simple expectations for a spherically expanding envelope.
At the extreme velocities, the amplitude is approximately constant, 
as expected for 
the unresolved ``caps'' of blueshifted and redshifted emission, while at 
intermediate velocities, the fall-off of visibility amplitude with 
$(u,v)$ distance demonstrates that the emission is spatially extended and 
resolved. If the gas is expanding radially with approximately spherical
symmetry, then the narrow velocity interval around the central velocity
corresponds to a cross section through the envelope in the plane of the sky.

The central peak of CS J=14--13 emission distribution provides additional 
direct evidence that CS is ``parent'' molecule, in accord with the 
detection of ro-vibrational transitions in absorption in the infrared 
(Keady \& Ridgeway 1993), the detection of vibrationally excited CS emission
(Turner 1987, Lucas \& Guelin 1999,  Highberger et al. 2000), 
and interferometric imaging of CS J=2--1 emission (Lucas et al. 1995). 
The CS molecules must originate close to the stellar photosphere, and 
they are lost from the expanding envelope through reactions that 
build more complex species, perhaps aided by shocks as suggested by 
Willacy \& Cherchneff (1998).
The J=14-13 emission distribution shows no indication of the 
extended $\sim15''$ radius ring visible in the J=2-1 line, 
but emission on that size scale, if present, could not be detected 
on account of the small SMA field of view.

The reaction network for sulfur bearing carbon chains has been explored
by Millar, Flores \& Markwick (2001), who concluded that an initial 
CS abundance of $4\times10^{-6}$ is needed to match observations of 
C$_3$S and C$_5$S. Models where CS is not a parent species, in which the 
CS fractional abundance rises from $<10^{-11}$ at radius $10^{16}$~cm, 
or $4\farcs$5 at 150~pc, to a peak more than four orders of magnitude larger 
at radius $10^{17}$~cm, could not match observations of the longer carbon 
chains. The SMA observations provide a robust {\em lower limit} to the 
CS abundance within a radius of $\sim34$~R$_*$ ($2.25\times10^{15}$~cm) 
that supports the parent species scenario.  
Under the assumption of a thermalized level population, the
total CS column density is given by
\begin{equation}
N(CS) = \frac{3 k f}{8 \pi^3 B \mu^2}
        \frac{\exp{(hBJ_l(J_l+1)/kT)}}{J_l+1} 
        \frac{T + hB/3k}{1- \exp{(-h\nu/kT)}}
        \int \tau dv,
\end{equation}
where the molecular constants are $\mu=1.96$ Debye, $B=24.584$ GHz, and
the filling factor $f$ may be estimated as (e.g. Scoville et al. 1986)
\begin{equation}
f = \frac{T_R^*/\eta_c}{(h\nu/k)/(\exp{(h\nu/kT)-1})(1 - \exp{-\tau}) }.
\end{equation}
The lower limit to the line flux in a $\sim2''$ beam is 
$\sim 5000$~Jy~km~s$^{-1}$, or $\sim 1250$~K~km~s$^{-1}$ 
in brightness units.
Assuming small optical depth, $\tau<<1$, a kinetic temperature equal to the 
excitation of the upper level of the observed transition (246.8~K), which 
minimizes the column density required to explain the observed line flux, 
and a spherical volume with molecular hydrogen density equal to the 
critical density of the transition of $3.2\times10^8$~cm$^{-3}$, results in 
a lower limit to the CS fractional abundance of $\sim3.4\times10^{-9}$. 
The abundance close to the photosphere is likely substantially higher
than this estimate; Highberger et al. (2000) derive $3-7\times10^{-5}$ 
at a radius $\sim14$~R$_*$ ($9\times10^{14}$~cm) from their multi-transition 
single dish observations. Though the uncertainties are large, models 
such as those investigated by Millar et al. (2001) where the CS abundance 
rises from $<10^{-11}$ at radius $10^{16}$~cm are clearly in conflict with 
the high resolution images of the CS J=14--13 emission.

The SMA observations show directly that the CS molecules are formed close 
to the stellar photosphere.  Detailed radiative transfer models that 
account for the observed spatial distribution are needed for an accurate 
determination of the CS abundance in the inner envelope. The effect of 
pulsational shocks could be important in establishing the CS abundance as 
this species is injected into the wind, and 
careful modeling may provide insight and constraints on the shock chemistry. 

Imaging the CS J=14--13 line in the inner envelope of IRC+10216 offers
a first demonstration of the efficacy of the SMA 
in the semi-transparent 450~$\mu$m atmospheric window. 
Calibration at these high frequencies should become considerably easier 
as more receivers for these frequencies are deployed at the SMA,
the correlator bandwidth is increased to 2~GHz, and simultaneous 
operation of a lower frequency band enables phase transfer. 
Many more evolved stars, as well as other objects detected but 
with presently unresolved emission at short submillimeter wavelengths, 
are now accessible to imaging at the arcsecond scale.

\acknowledgements
We thank all of members of the SMA Team that made these observations 
possible.
We thank John Bieging and the referee Jose Cernicharo for very helpful 
comments.
We extend special thanks to those of Hawaiian ancestry on whose sacred 
mountain we are privileged to be guests.


\begin{deluxetable}{lc}
\tablecolumns{3}
\tablewidth{0pt}
\tablecaption{IRC+10216 Observational Parameters}
\tablehead{
\colhead{Parameter} & \colhead{Value} }
\startdata
Observations: & 2002 Dec 10 (3 antennas)\\
Min/Max baseline: & 14 to 25~meters \\
Pointing center (J2000): & 
  $\alpha=9^{h}47^{m}57\fs39$, $\delta=13^{h}16^{m}43\fs90$ \\
Calibrator (flux): & Callisto (56 Jy) \\
Primary beam HPBW: & $17''$ \\
Synthesized beam HPBW: 
   & $2\farcs3\times1\farcs5$ P.A. $28^{\circ} $ \\
K/Jy:                 & 0.76 \\
Spectral Line Correlator: & $4\times128$ channels \\
~~~usable bandwidth: & 328 MHz \\
~~~species/transition: & CS J=14--13 \\
~~~frequency:          & 685.435923 GHz\tablenotemark{a}  \\
~~~channel spacing:  & 0.36 km~s$^{-1}$ \\
r.m.s. (line images): & 8 Jy~beam$^{-1}$ \\
\enddata
\tablenotetext{a}{Gottlieb, Myers \& Thaddeus 2003}

\label{tab:obs}
\end{deluxetable}

\clearpage

\clearpage

\clearpage

\end{document}